\documentclass[10pt,journal,compsoc]{IEEEtran}
\usepackage{graphicx}
\usepackage{subfigure}
\usepackage{amsmath}
\usepackage{amssymb}
\usepackage{color}
\usepackage{epstopdf}
\ifCLASSOPTIONcompsoc
  \usepackage[nocompress]{cite}
\else
  \usepackage{cite}
\fi
\ifCLASSINFOpdf
\else
\fi
\begin{document}
%
\title{Edge corona product as an approach to modeling complex simplical networks}

\author{Yucheng Wang, Yuhao Yi, Wanyue Xu and~Zhongzhi~Zhang,~\IEEEmembership{Member,~IEEE}
\IEEEcompsocitemizethanks{\IEEEcompsocthanksitem
Yucheng Wang, Yuhao Yi, Wanyue Xu,  and Zhongzhi~Zhang are with the Shanghai Key Laboratory of Intelligent Information Processing, School of Computer Science, Fudan University, Shanghai, 200433, China; and  also with  Shanghai Blockchain Engineering Research Center, Shanghai 200433, China.  (Corresponding author: Zhongzhi~Zhang.)  \protect\\
(E-mail: 15307130038@fudan.edu.cn;   15110240008@fudan.edu.cn;   xuwy@fudan.edu.cn; zhangzz@fudan.edu.cn).}
}

\IEEEtitleabstractindextext{%
\begin{abstract}
Many graph products  have been applied to generate complex networks with striking properties observed in real-world systems.  In this paper, we propose a simple generative model for simplicial networks by iteratively using edge corona product. We present a comprehensive analysis of the structural properties of the network model,  including  degree distribution, diameter, clustering coefficient, as well as  distribution of clique sizes, obtaining explicit expressions for these  relevant quantities, which  agree  with the behaviors  found in diverse real networks. Moreover, we obtain  exact expressions for  all the eigenvalues and their associated multiplicities of the normalized Laplacian matrix, based on which we  derive explicit formulas for mixing time, mean hitting time  and the number  of spanning trees. Thus, as previous models generated by  other graph products, our model is also an exactly solvable one, whose structural  properties can be  analytically treated. More interestingly, the expressions for the spectra of our model are also exactly determined,  which is sharp contrast to  previous models whose spectra  can only be given recursively at most.  This advantage makes our model  a good test-bed and an ideal substrate network for studying  dynamical processes, especially those closely related to the spectra of  normalized Laplacian matrix, in order to uncover the influences of simplicial structure on these processes.
\end{abstract}

}

\maketitle

\IEEEdisplaynontitleabstractindextext

%
\IEEEpeerreviewmaketitle



%
%
%
%
%
%
%

\IEEEraisesectionheading{\section{Introduction}\label{sec:introduction}}

\IEEEPARstart{C}{omplex} networks are a powerful tool for describing and studying the behavior of structural and dynamical aspects of complex systems~\cite{Ne03}. An important achievement in the study of complex networks is the discovery that various real-world systems from biology to social networks display some universal topological features, such as scale-free behavior~\cite{BaAl99} and small-world effect~\cite{WaSt98}. The former implies that the fraction of vertices with degree $d$ obeys a distribution of power-law form $P(d)\sim d^{-\gamma}$ with $2 < \gamma \leq 3$. The latter is characterized by small average distance (or diameter) and high clustering coefficient~\cite{WaSt98}. In addition to these two topological aspects, a lot of real networks are abundant in nontrivial patterns, such as $q$-cliques~\cite{Ts15} and many cycles  at different scales~\cite{RoKiBoBe05,KlSt06}.  
For example, spiking neuron populations form cliques in neural networks~\cite{GiPaCuIt15,ReNoScet17}, while coauthors of a paper constitute a clique in scientific collaboration networks~\cite{PaPeVa17}.  These remarkable structural properties or patterns greatly affect combinatorial~\cite{ZhWu15,JiLiZh17}, structural~\cite{ChLu02} and dynamical~\cite{ChWaWaLeFa08,YiZhPa20} properties of networks, and lead to algorithmic efforts on finding nontrivial subgraphs, e.g., $q$-cliques~\cite{MiPaPeTsXu15,JaSe17}.

In order to capture or account for universal properties observed in practical networks, a lot of mechanisms, approaches, and models were developed in the community of network science~\cite{Ne03}. Currently, there are many important graph generation literature~\cite{ChZhFa04,ViLa05,VeVade16}, graph generators~\cite{BaMa06}, as well as packages~\cite{LoPoSuBaScPo13}, for example, NetwrokX\footnote{https://networkx.github.io/}. In recent years, cliques, also called simplicial  complexes,  have become very popular to model complex networks~\cite{MiPaPeTsXu15,PeBa18}.  Since large real-world networks are usually made up of small pieces, for example, cliques~\cite{Ts15}, motifs~\cite{MiShItKaChAl02}, and communities~\cite{GiNe02}, graph products are an important and natural way for modelling real networks, which generate a large graph out of two or more smaller ones. An obvious  advantage of graph operations is the allowance of tractable analysis on various  properties of the resultant composite graphs. In the past years, various graph products have been exploited to mimic real complex networks, including Cartesian product~\cite{ImKl00}, corona product~\cite{LvYiZh15,QiLiZh18}, hierarchical product~\cite{BaCoDaFi09,BaDaFiMi09,BaCoDaFi16,QiYiZh18}, and Kronecker product~\cite{We62,LeFa07,MaXu07,LeChKlFaGh10,MaXu11}, and many more~\cite{PaNgBoKnFaRo11}.

Most current models based on graph operations either fail to reproduce serval properties of real networks or are hard to exactly analyze their spectral properties. For example, iterated corona product on complete graphs only yields small cycles~\cite{LvYiZh15,QiLiZh18}; while for most networks created by graph products, their spectra can be determined recursively at most.  On the other hand, in many real networks~\cite{BeAbScJaKl18,SaCaDaLa18}, such as brain networks~\cite{GiPaCuIt15,ReNoScect17} and protein-protein interaction networks~\cite{WuOtBa03}, there exist higher-order nonpairwise relations between more than two nodes at a time. These higher-order interactions, also called simplicial interactions, play an important role in other structural and dynamical properties of networks, including percolation~\cite{BiZi18}, synchronization~\cite{SkAr19,MiToBi19}, disease spreading~\cite{LaPeBaLa19}, and voter~\cite{HoKu20}. Unfortunately, most models generated by graph products and generators cannot capture  higher-order interactions, and how  simplicial interactions affect random walk dynamics, i.e., mixing time~\cite{LePeWi08}, is still unknown.  

From a network perspective, higher-order interactions can be described and modelled by hypergraphs~\cite{KlHaTh09,GhViCaNe09}. Here we model the higher-order interactions by simplicial complexes~\cite{Ha02} generated by a graph product. Although both simplicial complexes and hypergraphs can be applied for the modelling and analysis of realistic systems with higher-order interactions, they differ in some aspects. First, simplicial complexes have a geometric interpretation~\cite{DeVa19}. For example, they can be explained as the result of gluing nodes, edges, triangles, tetrahedra, etc. along their faces. This interpretation for simplicial complexes can be exploited to characterize the resulting network geometry, such as network curvatures~\cite{Ol09}. Moreover, a higher-order interaction described by hypergraphs do not require the presence of all low-order interactions.

In this paper, by literately applying edge corona product~\cite{HoYa10} first proposed by Haynes and Lawson~\cite{HaLa93,HaLa95} to complete graphs or   $q$-cliques $\mathcal{K}_q$ with $q\geq 1$, we propose a mathematically tractable model for complex networks with various cycles at different scales. Since the resultant networks are composed of cliques of different sizes, we call these networks as \textit{simplical networks}. The networks can describe simplicial interactions, which have rich structural, spectral, and dynamical properties depending on the parameter $q$. Thus, they can be used to study the influence of simplicial interactions on various dynamics.

Specifically, we present an extensive and exact analysis of relevant topological properties for the simplical networks, including degree distribution, diameter, clustering coefficient, and distribution of clique sizes, which reproduce the common properties observed for real-life networks. We also determine exact expressions for all the eigenvalues and their multiplicities of the transition probability matrix and normalized Laplacian matrix. As applications, we further exploit the obtained eigenvalues to derive leading scaling for mixing time, as well as explicit expressions for average hitting time and the number of spanning trees. The proposed model allows for rigorous analysis of structural properties, as previous models generated by graph products. In contrast to previous models for which the eigenvalues for related matrices are given recursively at most, the eigenvalues of transition probability matrix for our model can be exactly determined. This advantage allows to study analytically even exactly related dynamical processes determined by one or several eigenvalues, for example, mixing time of random walks, which gives deep insight into behavior for mixing time in real-life networks.

\section{Network construction}

The network family  proposed and studied here is constructed based on the edge corona product of graphs defined as follows~\cite{HoYa10,HaLa93,HaLa95}, which is a variant of the corona product first introduced by Frucht and Harary~\cite{FrHa70} of two graphs.  Let $\mathcal{G}_{1}$ and $\mathcal{G}_{2}$ be two graphs with disjoint vertex sets, with the former  $\mathcal{G}_{1}$ having $n_1$ vertices and $m_1$ edges. The  edge corona $\mathcal{G}_{1}  \circledcirc \mathcal{G}_{2}$ of  $\mathcal{G}_{1}$ and $\mathcal{G}_{2}$  is a graph obtained by taking one copy of $\mathcal{G}_{1}$ and $m_1$ copies of $\mathcal{G}_{2}$, and
then connecting both end vertices of the $i$th edge of  $\mathcal{G}_{1}$ to each vertex in the $i$th copy
of  $\mathcal{G}_{2}$ for $i=1,2,\ldots, m_1$.

\begin{figure}[htb]
\begin{center}
\includegraphics[width=0.9\linewidth]{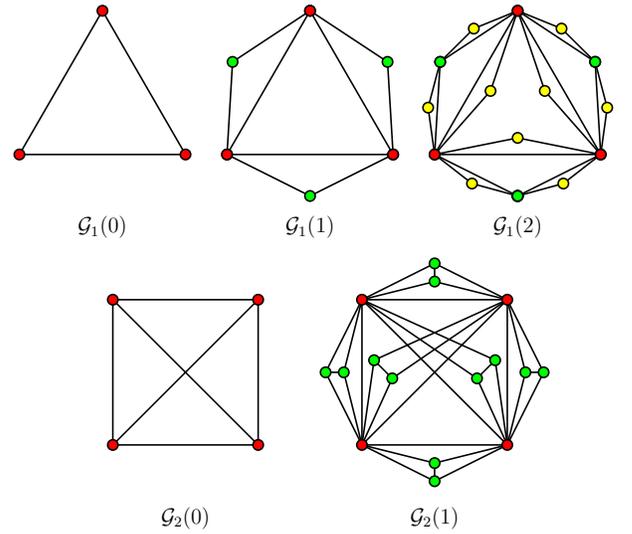}
\caption{The first several iterations of $\mathcal{G}_{q}(g)$ for   $q=1$ and $q=2$.}
\label{net-ex}
\end{center}
\end{figure}

Let $\mathcal{K}_q$, $q\geq 1$, be the  complete graph with  $q$ vertices. When $q= 1$, we define $\mathcal{K}_q$ as a graph with an isolate vertex. Based on the edge corona product and the  complete graphs, we can iteratively build a set of graphs, which display the  striking properties of real-world networks.  Let $\mathcal{G}_{q}(g)$, $q \geq 1$ and $g\geq 0$,  be  the network after  $g$  iterations.  Then, $\mathcal{G}_{q}(g)$  is constructed in the following way.
\newtheorem{definition}{Definition}
\begin{definition}\label{defa}
For $g=0$, $\mathcal{G}_{q}(0)$ is the complete graph $\mathcal{K}_{q+2}$. For $g\geq1$, $\mathcal{G}_{q}(g+1)$ is obtained from $\mathcal{G}_{q}(g)$ and $\mathcal{K}_{q}$ by performing  edge corona product on them: for every existing edge of $\mathcal{G}_{q}(g)$, we introduce a copy of the complete graph $\mathcal{K}_q$ and  connect  all its vertices to both end vertices of the edge.  That is,  $\mathcal{G}_{q}(g+1)= \mathcal{G}_{q}(g) \circledcirc \mathcal{K}_{q}$.
\end{definition}
Figure~\ref{net-ex}  illustrates the construction process of $\mathcal{G}_{q}(g)$ for two particular cases of  $q=1$ and $q=2$. Note that for   $q=1$, $\mathcal{G}_{q}(g)$ is reduced to the pseudofractal scale-free web~\cite{DoGoMe02}, which only contains triangles but excludes other complete graphs with more than 3 vertices.

Let $N_q(g)$ and $M_q(g)$ be the number of vertices  and number of edges in graph $\mathcal{G}_q(g)$, respectively. Suppose $L_v(g)$ and $L_e(g)$ be the number of vertices and the number of edges generated at iteration $g$. Then for $g=0$,  $L_v(0)=N_q(0)=q+2$ and $L_e(0)=M_q(0)=\frac{(q+1)(q+2)}{2}$. For all $g \geq 1$, by  Definition~\ref{defa},  we obtain the following two relations:
\begin{equation}\label{eqem}
L_v(g+1)=q M_q(g)
\end{equation}
and
\begin{equation}\label{eqmm}
L_e(g+1)=\left[\frac{(q+1)(q+2)}{2}-1\right] M_q(g),
\end{equation}
which lead to  recursive relationships for  $N_q(g)$ and $M_q(g)$ as
\begin{equation}
M_q(g+1)=\frac{(q+1)(q+2)}{2}M_q(g)
\end{equation}
and
\begin{equation}
N_q(g+1)=q M_q(g)+N_q(g).
\end{equation}
Considering the initial conditions $N_q(0)=q+2$ and $M_q(0)=\frac{(q+1)(q+2)}{2}$, the above two equations are solved to obtain
\begin{equation}\label{Mqg}
 M_q(g)={\left[\frac{(q+1)(q+2)}{2}\right]}^{g+1}
\end{equation}
and
\begin{equation}\label{Nqg}
N_q(g)=\frac{2}{q+3}{\left[\frac{(q+1)(q+2)}{2}\right]}^{g+1}+\frac{2(q+2)}{q+3}.
\end{equation}

Then, the average degree of vertices in graph $\mathcal{G}_q(g)$ is  $2M_q(g)/N_q(g)$, which  tends to $q+3$ when $g$ is large. Therefore,  the graph family $\mathcal{G}_q(g)$ is sparse.

In addition, inserting Eqs.~\eqref{Mqg} and~\eqref{Nqg} into Eqs.~\eqref{eqem} and~\eqref{eqmm} gives $L_v(g)=q \left[\frac{(q+1)(q+2)}{2}\right]^{g}$ and  $L_e(g)=\left[\frac{(q+1)(q+2)}{2}-1\right] \left[\frac{(q+1)(q+2)}{2}\right]^{g}$ for $g\geq1$, which are helpful for the computation in the sequel.

\section{Structural properties}

In this section, we study some relevant structural characteristics of $\mathcal{G}_q(g)$, focusing on degree distribution, diameter, clustering coefficient, and  distribution of clique sizes.

\subsection{Degree distribution}

The degree distribution $P(d)$ for a network is  the probability of a randomly selected vertex $v$ has exactly $d$ neighbors. When a network has a discrete sequence of vertex degrees, one can also use cumulative degree distribution $P_{\rm cum}(d)$ instead of ordinary degree distribution~\cite{Ne03}, which is the probability that a vertex has degree greater than or equal to $d$:
\begin{equation}\label{dd0}
P_{\rm cum}(d)= \sum_{d'=d}^{\infty}P(d').
\end{equation}
For a graph with degree distribution of power-law form $P(d)\sim d^{-\gamma}$, its cumulative degree distribution is also power-law satisfying $P_{\rm cum}(d)\sim d^{-(\gamma-1)}$.

For every vertex in graph $\mathcal{G}_q(g)$, its degree can be explicitly determined.  Let $d_v(g)$ be the degree of a vertex $v$ in graph $\mathcal{G}_q(g)$. When $v$ was generated at iteration $g_v$, it has a degree of $q+1$. By construction, for any edge incident with  $v$ at current iteration, it will lead to $q$ additional new edges adjacent to $v$ at the following iteration. Therefore,
\begin{equation}\label{degreeG}
d_v(g)=(q+1)^{g-g_v+1}\,.
\end{equation}
On the other hand,  in graph  $\mathcal{G}_q(g)$  the degree of all simultaneously emerging vertices is the
same.  Then, the number of vertices with the degree  $(q+1)^{g-g_v+1}$ is $q+2$ and $q \left[\frac{(q+1)(q+2)}{2}\right]^{g_v}$ for $g_v= 0$ and $g_v> 0$, respectively.
\newtheorem{proposition}{Proposition}
\begin{proposition}
The degree distribution of graph $\mathcal{G}_q(g)$ follows a power-law form $P(d)\sim d^{-\gamma}$ with the power exponent $\gamma=2+\frac{\ln(q+2)}{\ln(q+1)}-\frac{\ln 2}{\ln(q+1)}$.
\end{proposition}
\begin{IEEEproof}
As shown above, the degree sequence of vertices in $\mathcal{G}_q(g)$ is discrete.  Thus we can get the degree distribution $P(d)$ for $d=(q+1)^{g-g_v+1}$ via the cumulative degree distribution given by
\begin{align}
P_{\rm cum}(d)&=\frac{1}{N_q(g)}\sum_{\tau\leqslant g_v}L_v(\tau) \notag \\
&=\frac{\left[\frac{1}{2}(q+1)(q+2)\right]^{g_v+1}+q+2}{\left[\frac{1}{2}(q+1)(q+2)\right]^{g+1}+q+2}\,.
\end{align}
From Eq.~\eqref{degreeG}, we derive $g_v=g+1-\frac{\ln d}{\ln(q+1)}$. Plugging this expression for $g_v$ into the above equation leads to
\begin{equation}
\begin{split}
&P_{\rm cum}(d)=\frac{2^{\frac{\ln d}{\ln(q+1)}-g-2} \left[(q+1) (q+2)\right]^{-\frac{\ln d}{\ln (q+1)}+g+2}+q+2}{2^{-g-1}
   \left[(q+1) (q+2)\right]^{g+1}+q+2}\\
&=\frac{d^{-\left(\frac{\ln(q+2)}{\ln(q+1)}+1-\frac{\ln 2}{\ln(q+1)}\right)}  2^{-g-2}\left[(q+1) (q+2)\right]^{g+2}+q+2}{2^{-g-1}\left[(q+1) (q+2)\right]^{g+1}+q+2}.
\end{split}
\end{equation}
When $g\rightarrow \infty$, we obtain
\begin{equation}
P_{\rm cum}(d)=\frac{(q+1)(q+2)}{2}  d^{-\left(\frac{\ln(q+2)}{\ln(q+1)}+1-\frac{\ln 2}{\ln(q+1)}\right)}.
\end{equation}
So the degree distribution follows a power-law form $P(d)\sim d^{-\gamma}$ with the exponent $\gamma=2+\frac{\ln(q+2)}{\ln(q+1)}-\frac{\ln 2}{\ln(q+1)}$.
\end{IEEEproof}

It is not difficult  to see that the power exponent $\gamma$ lies in the interval $[\frac{\ln2}{\ln3}+2,3]$. Moreover, it is a monotonically increasing  function of $q$:  When $q$ increases from $2$ to infinite, $\gamma$ increases from $\frac{\ln 2}{\ln 3}+2$ to $3$. Note that for  most real scale-free networks~\cite{Ne03}, their power exponent $\gamma$ is in the  range between $2$ and $3$.

\subsection{Diameter}

In a graph $\mathcal{G}$, where every edge having unit length, a shortest path between a pair of vertices $u$ and $v$ is a path connecting $u$ and $v$ with least edges. The distance  $d(u,v)$ between $u$ and $v$ is defined as the number of edges in such a shortest path. The diameter of  graph $\mathcal{G}$, denoted by $D(\mathcal{G})$, is the  maximum of the  distances among all pairs of vertices.

\begin{figure}[htb]
\begin{center}
\includegraphics[width=.6\linewidth]{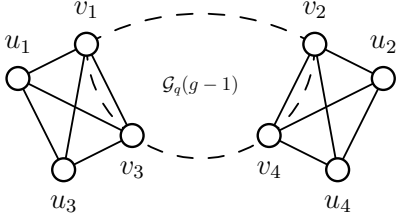}
\caption{Illustrative proof of the extended proposition.}
\label{diam2}
\end{center}
\end{figure}

\begin{proposition}
The diameter  $D(\mathcal{G}_q(g))$ of graph $\mathcal{G}_q(g)$, is $D(\mathcal{G}_1(g))=g+1$ for $q=1$ and $D(\mathcal{G}_q(g))=2g+1$ for $q\geq 2$.
\end{proposition}
\begin{IEEEproof}
For the case of $q=1$, $D(\mathcal{G}_1(g))=g+1$ was proved in~\cite{ZhRoZh07}. Below we only prove the case of $q\geq 2$.

For $g=0$, $D(\mathcal{G}_q(g))=1$, the statement holds. By Definition~\ref{defa}, it is obvious that the diameter of graph $\mathcal{G}_q(g)$ increases at most 2 after each iteration, which means  $D(\mathcal{G}_q(g))\leq 2g+1$. In order to prove $D(\mathcal{G}_q(g))=2g+1$, we only need to show that for $q\geq 2$ there exist two vertices in $\mathcal{G}_q(g)$, whose   distance $2g+1$. To this end,  we alternatively prove an extended proposition  that in $\mathcal{G}_q(g)$ there exist  two pairs of adjacent vertices: $u_1$ and $u_3$, $u_2$ and $u_4$, such that $d(u_1, u_2)=d(u_1, u_4)=d(u_3, u_2)=d(u_3, u_4)=2g+1$. We next prove this extended proposition by induction on $g$.

For $g=0$, $\mathcal{G}_q(0)$, $q\geq2$, is the complete graph $\mathcal{K}_{q+2}$.  We can arbitrarily choose four vertices  as $u_1$, $u_2$, $u_3$, $u_4$ to meet the condition. For $g \geq 1$, suppose that the statement holds for $\mathcal{G}_q(g-1)$, see Fig.~\ref{diam2}. In other words,   there exist two pairs of adjacent vertices: $v_1$ and $v_3$, $v_2$ and $v_4$ in $\mathcal{G}_q(g-1)$,  with their  distances  in  $\mathcal{G}_q(g-1)$ satisfying $d(v_1, v_2)=d(v_1, v_4)=d(v_3, v_2)=d(v_3, v_4)=2g-1$.  For $\mathcal{G}_q(g-1)$,  let   $u_1$ and $u_3$ be two adjacent vertices generated by the edge connecting $v_1$ and $v_3$ at iteration $g$, and  let   $u_2$ and $u_4$ be two adjacent vertices generated by the edge connecting $v_2$ and $v_4$ at iteration $g$. Then, by assumption, for the  vertex pair  $u_1$ and $u_2$   in graph $\mathcal{G}_q(g-1)$,  their distance obeys $d(u_1,u_2)=\min\{d(v_1, v_2), d(v_1, v_4), d(v_3, v_2), d(v_3, v_4)\}+2=2g+1$. Similarly, we can prove that in  $\mathcal{G}_q(g-1)$, the distances of related  vertex pairs  satisfy  $d(u_1, u_4)=d(u_3, u_2)=d(u_3, u_4)=2g+1$.
\end{IEEEproof}

From Eq.~\eqref{Nqg}, the number of vertices $N_q(g)\sim \left[\frac{(q+1)(q+2)}{2}\right]^{g+1}$. Thus,   the diameter $D(\mathcal{G}_q(g))$ of $\mathcal{G}_q(g)$ scales logarithmically with $N_q(g)$, which means that  the graph family $\mathcal{G}_q(g)$ is small-world.

\subsection{Clustering coefficient}

Clustering coefficient~\cite{WaSt98} is another crucial quantity characterizing network structure.
In a graph $\mathcal{G}=\mathcal{G}(\mathcal{V }, \mathcal{E})$ with vertex set $\mathcal{V }$ and edge set $\mathcal{E}$,  the clustering coefficient $C_v(\mathcal{G})$ of a vertex $v$ with degree $d_v$  is defined~\cite{WaSt98} as the ratio of the  number $\epsilon_v$ of edges between the neighbours of $v$ to the possible maximum value $d_v(d_v-1)/2$, that is $C_v(\mathcal{G})=\frac{2\epsilon_v}{d_v(d_v-1)}$.  The  clustering coefficient $C(\mathcal{G})$ of the whole network $\mathcal{G}$ is defined as the average of $C_v(\mathcal{G})$ over all vertices: $C(\mathcal{G})=\frac{1}{|\mathcal{V}|}\sum_{v\in \mathcal{V}}C_v(\mathcal{G})$.

For graph $\mathcal{G}_q(g)$,  the clustering coefficient for all vertices and their average value can be determined explicitly.
\begin{proposition}
\label{vClustering}
In graph $\mathcal{G}_q(g)$,  the clustering coefficient $C_v(\mathcal{G}_q(g))$ of any vertex with degree $d_v(g)$  is
\begin{equation}
C_v(\mathcal{G}_q(g))=\frac{q+1}{d_v(g)}\,.
\end{equation}
\end{proposition}
\begin{IEEEproof}
By Definition~\ref{defa},  when a vertex $v$ was created at iteration $g_{v}$, its degree and clustering coefficient are $q+1$ and 1, respectively.   In any  two successive  iterations $t$ and $t-1$ ($t \leq g$), its degrees   increases by a factor of $q$ as $d_v(t)=(q+1)d_v(t-1)$.  Moreover, once its degree increases by $q$, then the number of edges between its neighbors increases by $q(q+1)/2$. Then,  in network $\mathcal{G}_q(g)$, the clustering coefficient $C_v(\mathcal{G}_q(g))$ of vertex $v$ with degree
degree  $d_v(g)$ is
\begin{equation}
C_v(\mathcal{G}_q(g))=\frac{\frac{q(q+1)}{2}+\frac{d_v(g)-q-1}{q}\frac{q(q+1)}{2}}{\frac{d_v(g)(d_v(g)-1)}{2}}=\frac{q+1}{d_v(g)},
\end{equation}
as claimed by the Proposition.
\end{IEEEproof}

Thus, in graph  $\mathcal{G}_q(g)$, the  clustering coefficient of any vertex  is inversely proportional to its degree, a scaling  observed in various real-world  networked systems~\cite{RaBa03}.

\begin{proposition}
\label{gClustering}
For all $g\geq0$, the clustering coefficient of $\mathcal{G}_q(g)$ is
\begin{align}
&C(\mathcal{G}_q(g))=\nonumber\\
&\frac{{\left[\frac{(q+1)^2(q+2)}{2}\right]}^{g+1}+q^2+4q+4}{\frac{q^2+4q+5}{(q+1)(q+3)}{\left[\frac{(q+1)^2(q+2)}{2}\right]}^{g+1}+\frac{(q+2)(q^2+4q+5)}{q+3}{(q+1)}^{g}}\,.
\end{align}
\end{proposition}
\begin{IEEEproof}
By using Proposition~\ref{vClustering}, the quantity $C(\mathcal{G}_q(g))$ can be calculated by
\begin{small}
\begin{align}
&C(\mathcal{G}_q(g))=\frac{1}{N_q(g)}\Bigg(\sum_{g_v=0}^{g}L_v(g_v)\cdot\frac{q+1}{d_v(g)}\Bigg)\notag\\
&=\frac{1}{N_q(g)}\Bigg\{\frac{(q+2)}{(q+1)^{g}}+\sum_{g_v=1}^{g}q{\left[\frac{(q+1)(q+2)}{2}\right]}^{g_v} \frac{q+1}{(q+1)^{g-g_v+1}}\Bigg\}\notag\\
&=\frac{{\left[\frac{(q+1)^2(q+2)}{2}\right]}^{g+1}+q^2+4q+4}{\frac{q^2+4q+5}{(q+1)(q+3)}{\left[\frac{(q+1)^2(q+2)}{2}\right]}^{g+1}+\frac{(q+2)(q^2+4q+5)}{q+3}{(q+1)}^{g}}.
\end{align}
\end{small}
This finishes the proof.
\end{IEEEproof}

From Proposition~\ref{gClustering}, we can see that  the clustering coefficient of graph $\mathcal{G}_q(g)$ is very high.  For large $g$, the clustering coefficient $\mathcal{G}_q(g)$ converges to a large constant as
\begin{equation}
\lim_{g\rightarrow \infty}C(\mathcal{G}_q(g))= \frac{q^2+4q+3}{q^2+4q+5}\,.
\end{equation}
Thus, similarly to the degree exponent $\gamma$, clustering coefficient $C(\mathcal{G}_q(g))$ is also dependent on $q$, with large $q$ corresponding to large  $C(\mathcal{G}_q(g))$. When  $q \rightarrow \infty$, the clustering coefficient of the graph tends to $1$.

\subsection{Distribution of clique sizes}

It is apparent that graph $\mathcal{G}_q(g)$ contains many cliques as subgraphs. Let $N_k(G_q(g))$  denote the number of $k$-cliques in graph $\mathcal{G}_q(g)$. Since graph $\mathcal{G}_q(0)$ is a $q+2$ complete graph,  the maximum clique size in it is $q+2$. Then in  $\mathcal{G}_q(0)$ the number $N_k(\mathcal{G}_q(0))$ of $k$-cliques  is the combinatorial number $C_{q+2}^{k}=\frac{(q+2)!}{k!(q+2-k)!}$  for $k=2,3,\ldots,q+2$,  and is $0$ for $k>q+2$. For graph $\mathcal{G}_q(g)$ with $g\geq1$, the number of 2-cliques equals  the number of edges,  while for cliques with size more than 2, we have the following proposition.

\begin{proposition}\label{clique}
For $g\geq0$, we have
\begin{equation}
N_k(\mathcal{G}_q(g))=\frac{{\left[\frac{(q+1)(q+2)}{2}\right]}^{g+1}-1}{\frac{(q+1)(q+2)}{2}-1}\frac{(q+2)!}{k!(q+2-k)!},
\end{equation}
for $k=3,4,\ldots,q+2$. And $N_k(\mathcal{G}_q(g))=0$, for $k>q+2$.
\end{proposition}
\begin{IEEEproof}
The proposition is naturally satisfied in graph $\mathcal{G}_q(0)$. Thus, we only need to prove the proposition for $g\geq1$. By definition, when $g\geq1$,  $\mathcal{G}_q(g)$ is obtained from $\mathcal{G}_q(g-1)$ by introducing a new $q$-complete graph for every edge. Then, all the $k$-cliques in $\mathcal{G}_q(g)$ can be partitioned into two parts: (i) the $k$-cliques in  $\mathcal{G}_q(g-1)$, and (ii) the $k$-cliques that contain at least one newly  introduced vertex.

For part (i), the number of  $k$-cliques is $N_k(\mathcal{G}_q(g-1))$. For part (ii), since every  newly  introduced vertex is only connected to other vertices in the $q+2$ compete graph generated by an  edge of $\mathcal{G}_q(g-1)$, any $k$-clique contain this new vertex must be a subgraph of this $q+2$ compete graph. The number of new $q+2$ compete graphs equals  the number $M_q(g-1)$ of edges in $\mathcal{G}_q(g-1)$, and in every new  $q+2$ complete graph, the number of $k$-cliques  is the combinatorial number $C_{q+2}^{k}$ for $k\leq q+2$. Since in every new $q+2$ complete graph, there are only two old vertices,  each of its $k$-clique subgraph with $k\geq3$ includes at least one new vertex. Thus, for part (ii) the number of  $k$-cliques can be calculated by $M_q(g-1)C_{q+2}^{k}$ for $3\leq k\leq q+2$, and is obviously 0 for $k>q+2$.

Combining the above results,  we have that for $g\geq1$,
 \begin{equation}
N_k(\mathcal{G}_q(g))=N_k(\mathcal{G}_q(g-1))+M_q(g-1)C_{q+2}^{k},
 \end{equation}
 for $3\leq k\leq q+2$, and $N_k(\mathcal{G}_q(g))=N_k(\mathcal{G}_q(g-1))$ for $k>q+2$.
 Together with $M_q(g-1)={\left[\frac{(q+1)(q+2)}{2}\right]}^g$, $C_{q+2}^{k}=\frac{(q+2)!}{k!(q+2-k)!}$, and the initial values for $\mathcal{G}_q(0)$, the above recursive relation is  solved to obtain the proposition.
\end{IEEEproof}

\section{Spectra of  probability transition matrix and normalized Laplacian  matrix}

Let $\textbf{A}_g=\textbf{A}(\mathcal{G}_q(g))$ denote the  adjacency matrix of  graph $\mathcal{G}_q(g)$, the entries $A_g(i,j)$ of which are defined as follows: $A_g(i,j) =1$ if the vertex pair of $i$ and $j$ is adjacent in $\mathcal{G}_q(g)$ by an edge denoted by $i \sim j$, or $A_g(i,j) =0$ otherwise.  The vertex-edge  incident matrix $\textbf{R}_g=\textbf{R}(\mathcal{G}_q(g))$ of graph $\mathcal{G}_q(g)$ is an $N_q(g)\times M_q(g)$ matrix,  the entries $R_g(v,e)$ of which are defined in the following way:
 $R_g(v,e)=1$ if   vertex $v$ is incident to   edge $e$, and $R_g(v,e)=0$ otherwise.  The diagonal degree matrix of $\mathcal{G}_q(g)$ is $\textbf{D}_g=\textbf{D}(\mathcal{G}_q(g))={\rm diag} \{d_1(g),d_2(g),\ldots,d_{N_q(g)}(g)\}$, where the $i$th nonzero entry is the degree $d_i(g)$ of vertex $i$ in graph $\mathcal{G}_q(g)$.  The Laplacian matrix $\textbf{L}_g=\textbf{L}(\mathcal{G}_q(g))$ of graph  $\mathcal{G}_q(g)$ is  $\textbf{L}_g=\textbf{D}_g-\textbf{A}_g$. The transition probability matrix of $\mathcal{G}_q(g)$, denoted by $\textbf{P}_g=\textbf{P}(\mathcal{G}_q(g))$, is defined by $\textbf{P}_g=\textbf{D}_g^{-1}\textbf{A}_g$, with the $(i,j)$th element $P_g(i,j)=1/d_i(g)$ representing the  transition probability for a walker going from vertex $i$ to vertex $j$ in graph $\mathcal{G}_q(g)$. Matrix $\textbf{P}_g$ is  asymmetric, but is similar to the normalized adjacency matrix $\tilde{\textbf{A}}_g(\mathcal{G}_q(g))=\tilde{\textbf{A}}_g$ of   graph  $\mathcal{G}_q(g)$ defined by $\tilde{\textbf{A}}_g=\textbf{D}_g^{-\frac{1}{2}}\textbf{A}_g\textbf{D}_g^{-\frac{1}{2}}$, since $\tilde{\textbf{A}}_g=\textbf{D}_g^{-\frac{1}{2}}\textbf{P}_g\textbf{D}_g^{\frac{1}{2}}$.  By definition, the $(i,j)$th entry of matrix $\tilde{\textbf{A}}_g$ is $\tilde{A}_g(i,j) =\frac{A_g(i,j)}{\sqrt{d_i(g)}\sqrt{d_j(g)}}$. Thus, matrix $\tilde{\textbf{A}}_g$ is real and symmetric, and has the same set of eigenvalues as the transition probability matrix $\textbf{P}_g$.  For  graph  $\mathcal{G}_q(g)$, its normalized Laplacian matrix $\tilde{\textbf{L}}_g(\mathcal{G}_q(g))=\tilde{\textbf{L}}_g$ is defined by
$\tilde{\textbf{L}}_g=\textbf{I}_g-\tilde{\textbf{A}}_g$, where $\textbf{I}_g$ is the $N_q(g)\times N_q(g)$ identity matrix.

In the remainder of this section, we will study the full spectrum of transition probability matrix $\textbf{P}_g$ and normalized Laplacian matrix $\tilde{\textbf{L}}_g$ for graph $\mathcal{G}_q(g)$. For $i=1,2,\cdots, N_q(g)$, let $\lambda_i(g)=\lambda_i({G}_q(g))$ and $\sigma_i(g)=\sigma_i({G}_q(g))$ denote the $N_q(g)$ eigenvalues of matrices $\textbf{P}_g$ and $\tilde{\textbf{L}}_g$, respectively.  Let $\Lambda_g$ and $\Sigma_g$ denote the set of eigenvalues of matrices $\textbf{P}_g$ and $\tilde{\textbf{L}}_g$, respectively, that is $\Lambda_g=\{\lambda_1(g),\lambda_2(g),\ldots,\lambda_{N_q(g)}(g)  \}$ and $\Sigma_g=\{\sigma_1(g),\sigma_2(g),\ldots,\sigma_{N_q(g)}(g)  \}$.  It is obvious that for all $i=1,2,\cdots,N_q(g)$, the relation $\lambda_i(g)=1-\sigma_i(g)$ holds. Moreover, the eigenvalues of matrices $\textbf{P}_g$ and $\tilde{\textbf{L}}_g$ can be listed in a nonincreasing (or nondecreasing) order as: $1=\lambda_1(g) \ge \lambda_2(g) \ge \ldots \ge \lambda_{N_q(g)}(g)\geq -1$ and $0=\sigma_1(g) \le \sigma_2(g) \le \cdots \le \sigma_{N_q(g)}(g)\leq 2$.

The one-to-to correspondence $\lambda_i(g)=1-\sigma_i(g)$ between   $\lambda_i(g)$ and $\sigma_i(g)$, for all $i=1,2,\cdots,N_q(g)$,  indicates that if one  determines the eigenvalues of matrix $\textbf{P}_g$, then the eigenvalues of  matrix $\tilde{\textbf{L}}_g$  are  easily found.
\newtheorem{lemma}{Lemma}
\begin{lemma}\label{lemeig}
For $\lambda\neq-\frac{1}{q+1}$ and $\lambda\neq\frac{q-1}{q+1}$, $\lambda$ is an eigenvalue of $\textbf{P}_{g+1}$ if and only if $(q+1)\lambda-q$ is an eigenvalue of $\textbf{P}_g$, and the multiplicity of $\lambda$ of $\textbf{P}_{g+1}$, denoted by $m_{g+1}(\lambda)$, is the same as the multiplicity of eigenvalue $(q+1)\lambda-q$ of $\textbf{P}_g$, denoted by $m_{g}((q+1)\lambda-q)$, i.e. $m_{g+1}(\lambda)=m_{g}((q+1)\lambda-q)$.
\end{lemma}
\begin{IEEEproof}
Let $\mathcal{V}_{g+1}$ be the set of vertices in   graph $\mathcal{G}_q(g+1)$. It  can be looked upon the union of   two disjoint sets $\mathcal{V}_g$ and  $\mathcal{V'}_{g+1}=\mathcal{V}_{g+1}\backslash \mathcal{V}_{g}$, where $\mathcal{V'}_{g+1}$ includes all the newly introduced vertices by the edges in $\mathcal{G}_q(g)$. For all vertices in $\mathcal{V}_{g+1}$, we label those in $\mathcal{V}_{g}$ from 1 to $N_q(g)$, while label the vertices $\mathcal{V'}_{g+1}$ from $N_q(g)+1$ to $N_q(g+1)$. In the following statement, we represent all the vertices by their labels.

Let $\textbf{y}=(y_1,y_2,\ldots,y_{N_q(g+1)})^{\top}$ denote the eigenvector of eigenvalue $\lambda$ of matrix $\textbf{P}_{g+1}$, where the component $y_i$ corresponds to vertex $i$ in $\mathcal{G}_q(g+1)$. Then,
\begin{equation}\label{eqeig}
\lambda\,\textbf{y}=\textbf{P}_{g+1}\,\textbf{y}.
\end{equation}

By construction, for any two adjacent old vertices $u$ and $v$ in $\mathcal{V}_g$, there are $q$ vertices newly introduced by the edge connecting $u$ and $v$, which are denoted by $h_1$, $h_2$, $\ldots$, $h_q$.  These $q$ vertices, together with  $u$ and $v$ form a complete graph of $q+2$ vertices.  Moreover,  each  vertex $h_i$  in set $\{ h_1, h_2, \ldots, h_q\}$ is exactly connected to $u$, $v$, and other vertices in $\{ h_1, h_2, \ldots, h_q\}$ excluding $h_i$ itself. Then the row in Eq.~\eqref{eqeig} corresponding to vertex $h_i$, $i=1,2,\ldots,q$, can be written as
\begin{align}
\lambda\, y_{h_i}=&\sum_{j=1}^{N_q(g+1)}P_{g+1}(h_i,j)y_j\nonumber\\
=&\frac{1}{d_{h_i}(g+1)}\sum_{j\sim h_i}y_j\nonumber\\
=&\frac{1}{q+1}(y_u+y_v+y_{h_1}+\ldots +y_{h_{i-1}}\nonumber\\
&+y_{h_{i+1}}+\ldots+ y_{h_{q}})\,,
\end{align}
Adding $\frac{1}{q+1}y_{h_{i}}$ to both sides of the above equation yields
\begin{equation}\label{eqh1}
\left(\lambda+\frac{1}{q+1}\right)y_{h_i}=\frac{1}{q+1}\left(y_u+y_v+\sum_{j=1}^{q}y_{h_j}\right),
\end{equation}
for all $i=1,2,\ldots,q$. Therefore, for $\lambda\neq-\frac{1}{q+1}$,
\begin{equation}\label{eqh2}
y_{h_1}=y_{h_2}=\ldots=y_{h_q}.
\end{equation}
Combining Eqs.~\eqref{eqh1} and~\eqref{eqh2}, we can derive that, for $\lambda\neq\frac{q-1}{q+1}$
\begin{equation}\label{eqh3}
y_{h_i}=\frac{1}{(q+1)\lambda-q-1}(y_u+y_v)
\end{equation}
holds for $i=1,2,\ldots,q$.  According to Eq.~\eqref{eqeig}, we can also express the  rows corresponding to components $y_u$  and $y_v$. For the row associated with component $y_u$, we have
\begin{align}
\lambda\, y_{u}&=\sum_{j=1}^{N_q(g+1)} P _{g+1}(u,j)y_j\nonumber\\
&=\frac{1}{d_u(g+1)}\left( \sum_{\substack{ j\leq N_q(g) \\  j\sim u} }  y_j+\sum_{\substack{ j > N_q(g) \\  j\sim u} }y_j\right).\label{eqh4}
\end{align}
By Definition~\ref{defa}, for an old vertex $u$, all its adjacent vertices in $\mathcal{V'}_{g+1}$ are introduced by the edges between $u$ and its neighboring vertices in  $\mathcal{V}_g$. Thus, combining Eqs.~\eqref{eqh3} and~\eqref{eqh4}, we  derive
\begin{equation}\label{eqh5}
\lambda \, y_{u}=\frac{1}{d_u(g+1)}\left(\sum_{\substack{ j\leq N_q(g) \\  j\sim u} }y_j+\sum_{\substack{ j\leq N_q(g) \\  j\sim u} }\frac{q(y_u+y_j)}{(q+1)\lambda-q-1}\right).
\end{equation}
Considering  $d_u(g+1)=(q+1)d_u(g)$,  Eq.~\eqref{eqh5} can be recast as
\begin{align}
&\quad \left((q+1)\lambda-\frac{q}{{(q+1)}\lambda-q-1}\right) y_{u}\nonumber\\
&=\frac{1}{d_u(g)}\sum_{\substack{ j\leq N_q(g) \\  j\sim u} }\left(1+\frac{q}{{(q+1)}\lambda-q-1}\right) y_j.
\end{align}
When $\lambda\neq-\frac{1}{q+1}$ and $\lambda\neq\frac{q-1}{q+1}$, the above equation is simplified as
\begin{align}
\left[(q+1)\lambda-q\right]y_{u}&=\frac{1}{d_u(g)}\sum_{\substack{ j\leq N_q(g) \\  j\sim u} }y_j.\nonumber\\
&=\sum_{j=1}^{N_q(g)}P_g(u,j)y_j,\label{eqh6}
\end{align}
which implies  if $\textbf{y}=(y_1,y_2,\ldots,y_{ N_q(g)},\ldots,y_{ N_q(g+1)})^{\top}$ is an eigenvector of matrix $\textbf{P}_{g+1}$ associated with eigenvalue $\lambda$, then $\tilde{\textbf{y}}=(y_1,y_2,\ldots,y_{ N_q(g)})^{\top}$ is an eigenvector of matrix $\textbf{P}_{g}$ associated with eigenvalue $(q+1)\lambda-q$.

On the other hand, suppose that $\tilde{\textbf{y}}=(y_1,y_2,\ldots,y_{ N_q(g)})^{\top}$ is an eigenvector of matrix  $\textbf{P}_{g}$ associated with eigenvalue $(q+1)\lambda-q$, then $\textbf{y}=(y_1,y_2,\ldots,y_{},\ldots,y_{N_q(g+1)})^{\top}$ is an eigenvector of  matrix $\textbf{P}_{g+1}$ associated with eigenvalue $\lambda$ if and only if its components $y_i$, $i=N_q(g)+1$, $N_q(g)+2$, $\ldots$, $N_q(g+1)$, can be expressed by Eq.~\eqref{eqh3}. Thus, the number of linearly independent eigenvectors of $\lambda$ is the same as that of $(q+1)\lambda-q$. Since both  $\textbf{P}_{g}$ and $\textbf{P}_{g+1}$ are  normal matrices, which are diagonalizable,  the multiplicity of $\lambda$  (or $(q+1)\lambda-q$) is equal to the number of its linearly independent eigenvectors. Hence, $m_{g+1}(\lambda)=m_{g}((q+1)\lambda-q)$.
\end{IEEEproof}

Lemma~\ref{lemeig} indicates that except $\lambda \ne -\frac{1}{q+1}$ and $\frac{q-1}{q+1}$, all eigenvalues $\lambda$ of matrix $\textbf{P}_{g+1}$ can be derived from those of matrix $\textbf{P}_{g}$. However, it is easy to check that both $-\frac{1}{q+1}$ and $\frac{q-1}{q+1}$ are eigenvalues of matrix $\textbf{P}_{g+1}$. Moreover, their multiplicities can be determined explicitly.  The following lemma gives the multiplicity of $-\frac{1}{q+1}$, while  the multiplicity of $\frac{q-1}{q+1}$ will be provided later.
\begin{lemma}\label{lemeig2}
The multiplicity of $-\frac{1}{q+1}$ as an eigenvalue of matrix $\textbf{P}_{g+1}$ is $(q-1)M_q(g)+N_q(g)$, i.e. $m_{g+1}(-\frac{1}{q+1})=(q-1)M_q(g)+N_q(g)$.
\end{lemma}
\begin{IEEEproof}
Let $\textbf{y}=(y_1,y_2,\ldots\ldots,y_{N_q(g+1)})^{\top}$   be an eigenvector associated with eigenvalue $-\frac{1}{q+1}$ of matrix $\textbf{P}_{g+1}$. Then,
\begin{equation}\label{eqh7}
-\frac{1}{q+1}\,\textbf{y}=\textbf{P}_{g+1}\,\textbf{y}.
\end{equation}
For an edge $e_x$, $x=1,2\ldots, M_q(g)$, in graph $\mathcal{G}_q(g)$ with end vertices  $u$ and  $v$, at iteration $g+1$, it will generate $q$ vertices $h_1$, $h_2$, $\ldots$, $h_q$  in $\mathcal{V}' _{g+1}$.  Then,  the row in Eq.~\eqref{eqh7} corresponding to vertex $h_i$, $i=1,2,\ldots,q$, can be expressed by
\begin{align}
-\frac{1}{q+1}\,y_{h_i}=&\sum_{j=1}^{N_q(g+1)} P_{g+1}(h_i,j)\, y_j\nonumber\\
=&\frac{1}{q+1}(y_u+y_v+y_{h_1}+\ldots +y_{h_{i-1}}\nonumber\\
&+y_{h_{i+1}}+\ldots +y_{h_{q}}),
\end{align}
which is equivalent to
\begin{equation}\label{eqh8}
\sum_{i=1}^{q}y_{h_i}=-(y_u+y_v).
\end{equation}

On the other hand, the row in Eq.~\eqref{eqh7} corresponding to vertex ${u}$ can be expressed as
\begin{equation}
-\frac{1}{q+1}y_{u}=\frac{1}{d_u(g+1)}\left(\sum_{\substack{ j\leq N_q(g) \\  j\sim u} }y_j+\sum_{\substack{ j > N_q(g) \\  j\sim u} }y_j\right).\label{eqh9}
\end{equation}
Note that Eq.~\eqref{eqh8} holds for every pair of adjacent vertices  in graph $\mathcal{G}_q(g)$ and the $q$ new vertices it generates at iteration $g+1$. Plugging Eq.~\eqref{eqh8} into the right-hand side of Eq.~\eqref{eqh9} leads to
\begin{align}
&\quad \frac{1}{d_u(g+1)}\left(\sum_{\substack{ j\leq N_q(g) \\  j\sim u} }y_j+\sum_{\substack{ j > N_q(g) \\  j\sim u} }y_j\right) \notag \\
&=\frac{1}{d_u(g+1)}\left(\sum_{\substack{ j\leq N_q(g) \\  j\sim u} }y_j+\sum_{\substack{ j\leq N_q(g) \\  j\sim u} }-(y_u+y_j)\right)\nonumber\\
&=\frac{1}{d_u(g+1)}\left(\sum_{\substack{ j\leq N_q(g) \\  j\sim u} }-y_u\right)\nonumber\\
&=-\frac{1}{(q+1)}y_u.
\end{align}
Therefore,  the constraint on $\textbf{y}$ in Eq.~\eqref{eqh7} is equivalent to the constraint provided by $M_q(g)$ equations in Eq.~\eqref{eqh8}. The matrix form of these $M_q(g)$ equations can be written as
\begin{equation}\label{eqh10}
\setlength{\arraycolsep}{1.0pt}
\left[\begin{array}{cccc|cccccccccccccc}
& & & &  1&1&\cdots & 1& & & & & & &&&& \\
& & & &  && & &1 &1 &\cdots &1 & & &&&& \\
& &-\textbf{R}_g^{\top}  & & & & & & &&&&\ddots& & & & &  \\
& & & & & & & & & &&&&\cdots & & & &  \\
& & & & & & & &  &&&&& &1 &1 & \cdots & 1\\
\end{array}
\right]\textbf{y}=\textbf{0},
\end{equation}
where $\textbf{R}_g^{\top}$ is the transpose of $\textbf{R}_g$, and  the unmarked entries are vanishing. It is straightforward that the right partition of the matrix in Eq.~\eqref{eqh10} is an $M_q(g)\times qM_q(g)$ matrix, with each row corresponding to an edge $e_x$, $x=1,2,\ldots, M_q(g)$, in graph $\mathcal{G}_q(g)$. Moreover, in each row associated with $e_x$, $1$ repeats $q$ times, corresponding to the $q$ vertices newly created by  edge $e_x$.

Since the row vectors of the matrix in Eq.~\eqref{eqh10} are linearly independent, the dimension of the solution space of Eq.~\eqref{eqh10} is $N_q(g+1)- M_q(g)=(q-1) M_q(g)+ N_q(g)$. Therefore, the multiplicity of eigenvalue $-\frac{1}{q+1}$ for matrix  $\textbf{P}_{g+1}$  is $(q-1) M_q(g)+ N_q(g)$.
\end{IEEEproof}

\newtheorem{theorem}{Theorem}
\begin{theorem}\label{thP}
 Let $\Lambda_g$, $g \geq 0$, be the set of the $N_q(g)$ eigenvalues $\lambda_1(g)$, $\lambda_2(g)$, $\ldots$, $\lambda_{N_q(g)}(g)$ for matrix $\textbf{P}_g$, satisfying $1=\lambda_1(g) \ge \lambda_2(g) \ge \ldots \ge \lambda_{N_q(g)}(g)\geq -1$.   Then the $N_q(g+1)$ eigenvalues for  $\textbf{P}_{g+1}$ forming the set  $\Lambda_{g+1}$ can be listed in a descending order as
 \begin{align}
\Lambda_{g+1}=&\bigg\{\frac{\lambda_1(g)+q}{q+1},\frac{\lambda_2(g)+q}{q+1},\ldots,\frac{\lambda_{N_q(g)}(g)+q}{q+1},\nonumber\\
&\underbrace{\frac{q-1}{q+1},\frac{q-1}{q+1},\ldots,\frac{q-1}{q+1}}_{M_q(g)-N_q(g)},\nonumber\\
&\underbrace{-\frac{1}{q+1},-\frac{1}{q+1},\ldots,-\frac{1}{q+1}}_{(q-1)M_q(g)+N_q(g)}\bigg\}.  \label{XYZ}
\end{align}
\end{theorem}
\begin{IEEEproof}
We prove this theorem by induction on $g$.  First, for $g=0$, it is easy to verify that the statement holds.  For graph $\mathcal{G}_q(g)$, $g\geq1$, assume that the  relation between $\Lambda_{g-1}$ and $\Lambda_{g}$ is valid. We now prove that the result is true for graph $\mathcal{G}_q(g+1)$.

For each eigenvalue $\lambda_{i}(g) \in \Lambda_{g}$, $i=1,2,\ldots, N_q(g)$,  we have $\lambda_{i}(g)>-1$ by the assumption. Therefore,  for $i=1,2,\ldots, N_q(g)$,
\begin{equation}\label{eqrc}
\frac{\lambda_{i}(g)+q}{q+1}>\frac{q-1}{q+1},
\end{equation}
which implies  $\frac{\lambda_{i}(g)+q}{q+1}\neq \frac{q-1}{q+1}$ and $\frac{\lambda_{i}(g)+q}{q+1}\neq -\frac{1}{q+1}$. By Lemma~\ref{lemeig}, $\frac{\lambda_{i}(g)+q}{q+1}$ is an eigenvalue of $\textbf{P}_{g+1}$ with the same multiplicity of $\lambda_{i}(g)$ as an eigenvalue of $\textbf{P}_{g}$, namely,
\begin{equation}\label{eqrc2}
m_{g+1}\left(\frac{\lambda_{i}(g)+q}{q+1}\right)=m_{g}\left(\lambda_{i}(g)\right).
\end{equation}
Moreover, by Lemma~\ref{lemeig},  for each eigenvalue $\lambda$  of $\textbf{P}_{g+1}$ satisfying $\lambda\neq-\frac{1}{q+1}$ and $\lambda\neq\frac{q-1}{q+1}$,   $(q+1)\lambda-q$ must be an  eigenvalue of  $\textbf{P}_{g}$, which means $\lambda$ can be expressed by as $\lambda=\frac{\lambda_{i}(g)+q}{q+1}$ with $i \in \{1,2,\ldots,N_q(g) \}$. Therefore, the sum of multiplicity of all eigenvalues of $\textbf{P}_{g+1}$ excluding $-\frac{1}{q+1}$ and $\frac{q-1}{q+1}$ is $N_q(g)$, that is,
\begin{equation}\label{eqlm}
m_{g+1}\left(\lambda\notin\left\{-\frac{1}{q+1},\frac{q-1}{q+1}\right\}\right)=N_q(g).
\end{equation}

We proceed to compute the multiplicity $m_{g+1}\left(\frac{q-1}{q+1}\right)$ of eigenvalue $\frac{q-1}{q+1}$  for matrix $\textbf{P}_{g+1}$, which obeys
\begin{align}
&m_{g+1}\left(-\frac{1}{q+1}\right)+m_{g+1}\left(\frac{q-1}{q+1}\right)\nonumber\\
&+m_{g+1}\left(\lambda\notin\left\{-\frac{1}{q+1},\frac{q-1}{q+1}\right\}\right)=N_q(g+1).\label{eqeig2}
\end{align}
Using Eq.~\eqref{eqlm} and Lemma~\ref{lemeig2}, one obtains
\begin{equation}\label{eqeig3}
m_{g+1}\left(\frac{q-1}{q+1}\right)=M_q(g)-N_q(g).
\end{equation}
Combining Eqs.~\eqref{eqrc}~\eqref{eqrc2}~\eqref{eqeig3} and Lemma~\ref{lemeig2} yields~\eqref{XYZ}. 
\end{IEEEproof}

For $g=0$,  $\mathcal{G}_q(0)$ is a complete graph with $q+2$ vertices. The set of the eigenvalues of matrix $\textbf{P}_{0}$ is $\Lambda_{0}=\left\{1,-\frac{1}{q+1},-\frac{1}{q+1},\ldots,-\frac{1}{q+1}\right\}$. 
By recursively applying  Theorem~\ref{thP}, we can obtain all the eigenvalues matrix $\textbf{P}_g$ for   $g\geq 1$.



Using Theorem~\ref{thP} and the one-to-one correspondence between matrices $\tilde{\textbf{L}}_g$ and $\tilde{\textbf{P}}_g$, we can also obtain relation for  the set of eigenvalues for $\tilde{\textbf{L}}_g$ and $\tilde{\textbf{L}}_{g+1}$.

\begin{theorem}\label{thS}
 Let $\Sigma_g$, $g \geq 0$, be the set of the $N_q(g)$ eigenvalues $\sigma_1(g)$, $\sigma_2(g)$, $\ldots$, $\sigma_{N_q(g)}(g)$ for matrix $\tilde{\textbf{L}}_g$, satisfying $0=\sigma_1(g) \leq \sigma_2(g) \leq \ldots \leq \sigma_{N_q(g)}(g)\leq 2$.   Then the $N_q(g+1)$ eigenvalues for  $\tilde{\textbf{L}}_{g+1}$ forming the set  $\Sigma_{g+1}$ can be listed in an increasing order as
 \begin{align}
\Sigma_{g+1}&=\bigg\{\frac{\sigma_1(g)}{q+1},\frac{\sigma_2(g)}{q+1},\ldots,\frac{\sigma_{N_q(g)}(g)}{q+1},\nonumber\\
&\underbrace{\frac{2}{q+1},\frac{2}{q+1},\ldots,\frac{2}{q+1}}_{M_q(g)-N_q(g)},\underbrace{\frac{q+2}{q+1},\frac{q+2}{q+1},\ldots,\frac{q+2}{q+1}}_{(q-1)M_q(g)+N_q(g)}\bigg\}.
\end{align}
\end{theorem}
\begin{IEEEproof}
The proof is easily obtained by combining the relation $\lambda_i(g)=1-\sigma_i(g)$ and Theorem~\ref{thP}.
\end{IEEEproof}

The set $\Sigma_{0}$ of eigenvalues for matrix   $\tilde{\textbf{L}}_0$ is  $\Sigma_{0}=\left\{0,\frac{q+2}{q+1},\frac{q+2}{q+1},\ldots,\frac{q+2}{q+1}\right\}$.  
For $g\geq1$, by recursively applying Theorem~\ref{thS}, we can obtain the exact expressions for all eigenvalues for matrix $\tilde{\textbf{L}}_g$ for any $q$ and $g$, given by
\begin{align}\label{eqS2}
\Sigma_{g}=&\bigg\{0,\underbrace{\frac{q+2}{{(q+1)}^{g+1}},\frac{q+2}{{(q+1)}^{g+1}},\ldots,\frac{q+2}{{(q+1)}^{g+1}}}_{q+1},\notag\\
&\underbrace{\frac{2}{{(q+1)}^g},\frac{2}{{(q+1)}^g},\ldots,\frac{2}{{(q+1)}^g}}_{M_q(0)-N_q(0)},\notag\\
&\underbrace{\frac{q+2}{{(q+1)}^g},\frac{q+2}{{(q+1)}^g},\ldots,\frac{q+2}{{(q+1)}^g}}_{(q-1)M_q(0)+N_q(0)},\notag\\
&\underbrace{\frac{2}{{(q+1)}^{g-1}},\frac{2}{{(q+1)}^{g-1}},\ldots,\frac{2}{{(q+1)}^{g-1}}}_{M_q(1)-N_q(1)},\notag\\ &\underbrace{\frac{q+2}{{(q+1)}^{g-1}},\frac{q+2}{{(q+1)}^{g-1}},\ldots,\frac{q+2}{{(q+1)}^{g-1}}}_{(q-1)M_q(1)+N_q(1)},\notag\\ &\cdots \cdots,\notag\\
&\underbrace{\frac{2}{q+1},\frac{2}{q+1},\ldots,\frac{2}{q+1}}_{M_q(g-1)-N_q(g-1)},\underbrace{\frac{q+2}{q+1},\frac{q+2}{q+1},\ldots,\frac{q+2}{q+1}}_{(q-1)M_q(g-1)+N_q(g-1)}\bigg\}.
\end{align}

\section{Applications of the spectra}

In this section, we  apply the above-obtained  eigenvalues and their multiplicities  of related   matrices  to evaluate some relevant quantities for graph $\mathcal{G}_q(g)$, including mixing time, mean hitting time also called Kemeny constant,  and the number of spanning trees.

\subsection{Mixing time}

As is well-known, the probability transition matrix $\textbf{P}(\mathcal{G})$ of a graph $\mathcal{G}$ characterizes the process of   random walks  on the graph. As a classical Markov chain,   random walks describe various phenomena or  other  dynamical processes in graphs. Many interesting quantities about random walks can be extracted from the eigenvalues of the probability transition matrix. In this paper, we only consider mixing time and mean hitting time.

For  an ergodic random walk on an un-bipartite graph  $\mathcal{G}$ with $N$ vertices, it has a unique stationary distribution $\pi=(\pi_1, \pi_2,\ldots, \pi_N)^\top$ with  $\sum_{i=1}^{N}\pi_i=1$, where $\pi_i$ represents the probability that  the walker is at vertex $i$ when the random walk converges to equilibrium state~\cite{KeSn76}. The mixing time is defined as the expected time that the walker needs to approach the stationary distribution. Let $1=\lambda_1> \lambda_2 \geq \lambda_3 \geq \cdots \geq \lambda_N >-1$ be the $N$ eigenvalues for  matrix $\textbf{P}(\mathcal{G})$. Then the speed of convergence to the stationary distribution~\cite{Si92} approximately equals the reciprocal of $1-\lambda_{\max}$, where $\lambda_{\max}$ is the second largest eigenvalue modulus defined by $\lambda_{\max}=\max (\lambda_2, |\lambda_N|)$. Mixing time has found numerous applications in man different aspects~\cite{LePeWi08}.

As our first application of eigenvalues for matrix $\textbf{P}_g$, we use them to evaluate the mixing time for  random walks on $\mathcal{G}_q(g)$, for which the component of  stationary distribution $\pi$ corresponding to vertex $i$ is $\pi_i=d_i(g)/(2M_q(g))$. According to the above arguments, the  second largest eigenvalue modulus $\lambda_{\max}(g)$ of $\textbf{P}_g$ is $\lambda_{\max}(g)=1-\frac{q+2}{(q+1)^{g+1}}$. Since the mixing time is characterized by a parameter, it cannot be exactly  determined~\cite{Si92}, but one can evaluate it by using the reciprocal of $\lambda_{\max}(g)$. Then, the dominating term of the mixing time for random walks on  $\mathcal{G}_q(g)$ is $(q+1)^{g+1}/(q+2)$, which scales sublinearly with the vertex number $N_q(g)$ as $(N_q(g))^{2/\theta(q)}$, where $\theta(q)=2/ \log_{(q+1)(q+2)/2} (q+1)$ is the spectral dimension~\cite{MiToBi19} of graph $\mathcal{G}_q(g)$ that is a function of $q$. Note that for  $q=1$, the spectral dimension $\theta(2)=2 \ln 3 / \ln 2$ reduces to the result obtained in~\cite{BoDo20}.

Note that it is  believed that real-world networks  are often fast mixing with their mixing time  at most $O(\log N)$, where $N$ is the number of vertices.  However, it was experimentally  reported   that the mixing time of some real-world social networks is much higher than anticipated~\cite{MoYuKi10}.  Our obtained sublinear scaling of mixing time on graph $\mathcal{G}$  supports this recent study, and  sheds lights on understanding the scalings of mixing time.

 \subsection{Mean hitting time}

Our second application for our obtained eigenvalues  is the mean hitting time. For a random walk on graph  $\mathcal{G}$, the hitting time  $H_{ij}$, also called first-passage time~\cite{Re01,NoRi04,CoBeTeVoKl07},  from vertex $i$ to vertex $j$,  is defined as the expected time taken by a walker starting from vertex $i$ to   reach vertex $j$ for the first time. The mean hitting time $H$, also known as the Kemeny constant, is defined as the expected time for a random walker going from a vertex $i$ to another vertex $j$ that is chosen  randomly from all vertices in $\mathcal{G}$ according to the stationary distribution~\cite{LaLoPa96,AlFi02}:
\begin{equation}\label{eqhit}
H=\sum_{j=2}^{n}\pi_j H_{ij}.
\end{equation}
Interestingly,  the quantity $H$ is independent of   the starting vertex $i$, and  can be expressed in terms of the $N-1$ nonzero eigenvalues $\sigma_i$, $i=2,3,\cdots,N$, of the normalized Laplacian matrix $\tilde{\textbf{L}}(\mathcal{G})$ for graph $\mathcal{G}$, given by~\cite{LaLoPa96,AlFi02}
\begin{equation}\label{eqH}
H=\sum_{i=2}^{N} \frac{1}{\sigma_i}.
\end{equation}
Mean hitting time can be applied  to measure the efficiency of user navigation through the World Wide Web~\cite{LeLo02} and the efficiency of robotic surveillance in network environments~\cite{PaAgBu15}.  We refer to the reader to~\cite{Hu14} for many other   applications of mean hitting time.

In this subsection, we use the eigenvalues  of the normalized Laplacian matrix for graph $\mathcal{G}_q(g)$   to compute the mean hitting time of $\mathcal{G}_q(g)$.
\begin{theorem}\label{thMTT}
Let $H_{q}(g)$ be the mean hitting time for random walk in $\mathcal{G}_q(g)$. Then, for all $g\geq0$,
\begin{align}
H_{q}(g)=&\left[\frac{{(q+1)}^2}{q+2}-\frac{3(q+1)}{2}\right]{(q+1)}^{g}\notag\\
    &+\frac{(q+1)(3q+7)}{2(q+3)}{\left[\frac{(q+1)(q+2)}{2}\right]}^{g}+\frac{q+1}{q+3}. \label{Strees}
\end{align}
\end{theorem}
\begin{IEEEproof}
By Theorem~\ref{thS} and Eq.~\eqref{eqH}, we have
 \begin{align}
&H_{q}(g+1)\nonumber\\
    &=\frac{q+1}{2}\left(M_q(g)-N_q(g)\right)\nonumber\\
    &\quad+\frac{q+1}{q+2}\left((q-1)M_q(g)+N_q(g)\right)+\sum_{i=2}^{N_q(g)}\frac{q+1}{\sigma_i(g)}\nonumber\\
    &=\frac{3q(q+1)}{2(q+2)}M_q(g)-\frac{q(q+1)}{2(q+2)}N_q(g)+(q+1)H_{q}(g),\label{hitrelate}
\end{align}
which can be rewritten as
\begin{align}
&H_{q}(g+1)-\frac{(q+1)(3q+7)}{2(q+3)}{\left[\frac{(q+1)(q+2)}{2}\right]}^{g+1}-\frac{q+1}{q+3}\notag\\
 &=(q+1)\bigg\{H_{q}(g)-\frac{(q+1)(3q+7)}{2(q+3)}{\left[\frac{(q+1)(q+2)}{2}\right]}^{g}\notag\\
 &\quad-\frac{q+1}{q+3}\bigg\}.\label{hittime}
\end{align}
With the initial condition $ H_{q}(0)=\frac{{(q+1)^2}}{(q+2)}$, Eq.~\eqref{hittime} is solved to obtain~\eqref{Strees}.
\end{IEEEproof}

Theorem~\ref{thMTT} shows  that for $g \to \infty $, the dependence of mean hitting time $H_{q}(g)$ on the number $N_q(g)$ of vertices in   graph  $\mathcal{G}_g(g)$ is $H_{q}(g) \sim N_q(g)$, which implies that the $H_{q}(g)$ behaves linearly with $N_q(g)$.

\subsection{The number of spanning trees}

A spanning tree of  an undirected graph  $\mathcal{G}=(\mathcal{V},\mathcal{E})$  with $N$ vertices is a subgraph of  $\mathcal{G}$,  which  is a tree including all   the $N$ vertices. Let  $\tau(\mathcal{G})$ denote the number of spanning trees in graph $\mathcal{G}$. It has been shown~\cite{ChFa97,ChZh07} that $\tau(\mathcal{G})$ can be expressed in terms of the $N-1$ non-zero eigenvalues for normalized Laplacian matrix of $\mathcal{G}$ and the degrees  of all  vertices in $\mathcal{G}$:
\begin{equation}\label{eqt}
\tau(\mathcal{G})=\frac{\prod_{i\in \mathcal{V}}d_i  \prod_{i=2}^{N} \sigma_i(\mathcal{G})}{\sum_{i \in \mathcal{V}}d_i}.
\end{equation}

The number of spanning trees is an important graph invariant.  In the sequel,
we will use the above-obtained eigenvalues to determine this invariant for graph $\mathcal{G}_q(g)$.
\begin{theorem}
Let $\tau_q(g)=\tau(\mathcal{G}_q(g))$ be the number of spanning trees in graph  $\mathcal{G}_q(g)$. Then, for all $g\geq0$,
\begin{align}
\tau_q(g)&=2^{\frac{2(q+1)}{q{(q+3)}^2}{\left[\frac{(q+1)(q+2)}{2}\right]}^{g+1}-\left(\frac{q+1}{q+3}\right)g-\frac{{(q+1)}^2(q+2)}{{q(q+3)}^2}}\notag\\
&\cdot{(q+2)}^{\frac{2(q^2+2q-1)}{q{(q+3)}^2}{\left[\frac{(q+1)(q+2)}{2}\right]}^{g+1}+\left(\frac{q+1}{q+3}\right)g+\frac{q^3+2q^2-q+2}{{q(q+3)}^2}}. \label{SPtrees}
\end{align}
\end{theorem}
\begin{IEEEproof}
First,  by Theorem~\ref{thS}, we derive the relation for the product of all the non-zero eigenvalues for normalized Laplacian matrix for graph $\mathcal{G}_q(g+1)$ and $\mathcal{G}_q(g)$:
\begin{align}
&\prod_{i=2}^{N_q(g+1)} \sigma_i(g+1) \notag\\
&={\left(\frac{2}{q+1}\right)}^{M_q(g)-N_q(g)}
{\left(\frac{q+2}{q+1}\right)}^{(q-1)M_q(g)+N_q(g)}\prod_{i=2}^{N_q(g)}\frac{\sigma_i(g)}{q+1}\notag\\
 &=\frac{{2^{M_q(g)-N_q(g)}(q+2)}^{(q-1)M_q(g)+N_q(g)}}{{(q+1)}^{qM_q(g)+N_q(g)-1}}\prod_{i=2}^{N_q(g)} \sigma_i(g). \label{eqb}
\end{align}

Second, we derive the relation be between the product of degrees of all vertices in $\mathcal{G}_q(g+1)$ and the  product of degrees of all vertices in $\mathcal{G}_q(g)$. For $\mathcal{G}_q(g+1)$, the degree of all the new vertices in $\mathcal{V}'_{g+1}$ that were generated at iteration $g+1$ is  $q+1$; while for each $i$ of  those old vertices in $\mathcal{V}_{g}$, we have $d_i(g+1)=(q+1)d_i(g)$. Then,
 \begin{align}
       \prod_{i\in \mathcal{V}_{g+1}}d_i(g+1)&=\prod_{i\in \mathcal{V}'_{g+1}}d_i(g+1)\prod_{i\in \mathcal{V}_{g}}d_i(g+1)\notag\\
       &={(q+1)}^{qM_q(g)}\prod_{i\in \mathcal{V}_{g}}(q+1)d_i(g)\notag\\
       &={(q+1)}^{qM_q(g)+N_q(g)}\prod_{i\in \mathcal{V}_{g}}d_i(g).\label{eqc}
    \end{align}

Finally, the sum of degrees of all vertices in $\mathcal{G}_q(g)$ is equal to $2M_q(g)$.  Then,  Combining Eqs.~\eqref{Mqg},~\eqref{eqt},~\eqref {eqb}, and~\eqref{eqc}, we obtain the following recursive relation for $\tau_q(g+1)$ and $\tau_q(g)$:
\begin{equation}
\tau_q(g+1)=2^{M_q(g)-N_q(g)+1}{(q+2)}^{(q-1)M_q(g)+N_q(g)-1}\tau_q(g).
\end{equation}
Considering the expressions for $M_q(g)$ and $N_q(g)$ in Eqs.~\eqref{Mqg} and~\eqref{Nqg}, we obtain
 \begin{align}\label{spanT}
  \tau_q(g+1)&=2^{\frac{q+1}{q+3}{\left[\frac{(q+1)(q+2)}{2}\right]}^{g+1}-\frac{q+1}{q+3}}\nonumber\\
  &\quad \times{(q+2)}^{\frac{q^2+2q-1}{q+3}{\left[\frac{(q+1)(q+2)}{2}\right]}^{g+1}+\frac{q+1}{q+3}}\tau_q(g).
\end{align}
With the initial condition $\tau_q(0)=\tau(\mathcal{K}_{q+2})={(q+2)}^q$, Eq.~\eqref{spanT} is solved to yield~\eqref{SPtrees}.
\end{IEEEproof}
\section{Conclusion}

For many graph products  of two graphs,  one can  analyze the structural and spectral properties of the resulting graph, expressing them in terms those corresponding the two graphs. Because of this  strong advantage, many authors have used graph products to generate realistic networks with cycles at different scales. In this paper, by iteratively using the edge corona product, we  proposed a minimal model for complex networks called simplicial networks, which can capture  group interactions in real networks, characterized by a parameter $q$. We then provided an extensive  analysis for relevant topological properties of the model, most of which are dependent on $q$. We  show  that the resulting  networks display some remarkable characteristics of real  networks, such as  non-trivial higher-order interaction, power-law distribution of  vertex degree,  small diameter,  and  high clustering coefficient.

Furthermore, we found exact expressions for all the eigenvalues and their multiplicities of the transition probability matrix and normalized Laplacian matrix of our  proposed  networks. Using these obtained eigenvalues, we  further evaluated mixing time, as well as  mean hitting time for random walks on the  networks. The former scales sublinearly with the vertex number, while the latter behaves linearly with the vertex number. The sublinear scaling of mixing time is contrary to previous knowledge that  mixing time scales at most logarithmically with  the vertex number.  We also using the obtained eigenvalues to determine the number of spanning tree in the  networks.  Thus, in addition to the advantage of networks generated by other graph products, the proposed networks have another obvious advantage that both the eigenvalues and their multiplicities of relevant matrix can be analytically and exactly determined, since  for previous networks created by graph products,  the eigenvalues are only obtained recursively at most.  The explicit expression for each  eigenvalue  facilitates  to study those dynamical processes determined by one or several particular eigenvalues, such as mixing time considered here. 

It should be mentioned that many real networks are weighted with variable edge length~\cite{HeMaKoSw15}. For example, in scientific collaboration networks, the collaboration strength  between collaborators can be weighted by the number of papers they coauthored. It is thus necessary to model these realistic networks by weighted simplicial complexes~\cite{CoBi17}. In future, as the case of corona product~\cite{QiLiZh18}, one can also define extended  edge corona product of graphs and use it to build weighted scale-free networks with rich properties matching those of real-world networks~\cite{BaBaPaVe04}.





%



\ifCLASSOPTIONcompsoc
  \section*{Acknowledgments}
\else
  \section*{Acknowledgment}
\fi

This work was supported  in part by the National Natural Science Foundation of China (Nos. 61872093 and 61803248), the National Key R \& D Program of China
(No. 2018YFB1305104), Shanghai Municipal Science and Technology Major Project  (No.  2018SHZDZX01) and ZJLab.

\ifCLASSOPTIONcaptionsoff
  \newpage
\fi



%


\bibliographystyle{IEEEtran}
\bibliography{weight,Diagnol}

\end{document}